\documentclass[article,12pt,superscriptaddress,showkeys,showpacs] {revtex4}
\usepackage{amsfonts}
\usepackage{graphicx, amsmath}
\begin{document}

\title[Short Title]{Deterministic CNOT gate and entanglement swapping for photonic qubits using a quantum-dot spin in a double-sided optical microcavity}
\author{Hong-Fu Wang\footnote{E-mail: hfwang@ybu.edu.cn}}
\affiliation{Department of Physics, College of Science, Yanbian
University, Yanji, Jilin 133002, People's Republic of China}
\affiliation{School of Physics and Optoelectronic Technology, Dalian University of Technology, Dalian 116024,
People's Republic of China}
\author{Ai-Dong Zhu}
\affiliation{Department of Physics, College of Science, Yanbian
University, Yanji, Jilin 133002, People's Republic of China}
\author{Shou Zhang}
\affiliation{Department of Physics, College of Science, Yanbian
University, Yanji, Jilin 133002, People's Republic of China}
\author{Kyu-Hwang Yeon}
\affiliation{Department of Physics \& BK21 Program for Device
Physics, College of Natural Science, Chungbuk National University,
Cheongju, Chungbuk 361-763, Republic of Korea}
\begin{abstract}
We propose a deterministic and scalable scheme to construct a two-qubit controlled-NOT (CNOT) gate and realize entanglement swapping between photonic qubits using a quantum-dot (QD) spin in a double-sided optical microcavity. The scheme is based on spin selective photon reflection from the cavity and can be achieved in a nondestructive and heralded way. We assess the feasibility of the scheme and show that the scheme can work in both the weak coupling and the strong coupling regimes. The scheme opens promising perspectives for long-distance photonic quantum communication and distributed quantum information processing.
\end{abstract}

\maketitle \section{Introduction}\label{sec0}
Recent advances in cavity quantum electrodynamics (QED) has opened promising prospects in quantum computation and quantum state engineering. In cavity QED, the atoms act as the qubits and they are coupled via interacting with the cavity photon, schemes have been proposed to realize quantum computation and quantum communication\cite{ADAPRL9574, THPRL9574, PJMSPRA9552, TSJPPRL9575, LNMJSPRA9450, JAPRA9450, MPRA9654, XZCLCCPRA0775, HASKNJP1113}. On the other hand, the system of trapped ions, first proposed by Cirac and Zoller~\cite{JPPRL9574}, is an alternative qualified candidate for quantum information processing. Experimental reconstruction of the motional quantum state of a trapped ion has
been reported~\cite{DDBCWDPRL9677} and schemes for the generation of various motional states of a trapped ion~\cite{JRAPPRL9370, JARPPRL9370, RWPRL9676, JJRPPRL9654, CPRA9755, SPRA9858} and the implementation of quantum logic gates~\cite{CDBWDPRL9575, AKPRL9982, DMPRL0187, SPRL0390} have been proposed.
Furthermore, linear optical quantum computation (LOQC), which is recognized to be feasible following the demonstration~\cite{ERGN01409} that a scalable quantum computer can be built with linear optical elements, has attracted great interests in the past few years. The realization of a linear optical quantum computer is appealing due to the fact that photons are easily manipulated with high precision and, as the electro-magnetic environment at optical frequencies can be regarded as vacuum, are relatively decoherence free. Significant research effort has been dedicated in recent years to generating two-photon and multi-photon polarization entangled states~\cite{PKHAAYPRL9575, DIN99402,
XKWPRA04690266, GYPRA0265, HSPRA0979, XGPRA0878, YTMNPRA0571} and implementing quantum logic gates, such as controlled phase gate~\cite{HSPRA0266, NCURHPRL0595, XKGPRA0674, XSKGPRA0775, KAJKJMPRL11106, AGTMBPRA1286}, CNOT gate~\cite{TBJPRA0164, TNTAPRA0265, JGATDN03426, TMBJPRA0368, YXYLGPRL0493, ZAYHJTJPRL0594, XTQJHTJPRL0798, QJPRA0979, WPXOACCZJ10104}, SWAP gate~\cite{MTFPRA0572, AJLMJPRL08100}, Toffoli gate~\cite{JPRA0878}, Fredkin gate~\cite{JPRA0878, YGTPRA0878}, and so on. However, all these linear-optical quantum gates are probabilistic by their very nature~\cite{ERGN01409}, resulting in the fact that the probability for implementing universal quantum computation may be very tiny due to the complicated combination of so many nondeterministic two-qubit unitary gates. For example, the optimal probability of success for achieving a CNOT gate is $1/4$ with only single photon sources, a maximally entangled two-photon state, linear optical elements including feed-forward, photon detectors~\cite{TBJPRA0164}. It is well known that the SWAP gate can be constructed by three CNOT gates, or can be decomposed into six Hadamard gates and three controlled phase gates. So the probability of implementing a linear optical SWAP gate based on CNOT gate is $4^{-3}=1.56\times 10^{-2}$. While for Fredkin gate, which can be constructed by five CNOT gates and some single-qubit unitary gates, the probability is $4^{-5}=9.77\times 10^{-4}$. To overcome this inefficiency, much of the theoretical effort has focused on looking for more efficient ways to perform the controlled logic. Fortunately, assisted with the weak cross-Kerr nonlinearity, Nemoto and Munro~\cite{KWPRL0493} proposed an interesting scheme for constructing a nearly deterministic linear optical CNOT gate using homodyne-heterodyne measurement. With this architecture for the CNOT gate and feed-forward operations, a universal set of gates is hence possible for universal quantum computation in a deterministic way. However, it is not easy to implement the CNOT gate as proposed in Ref.~\cite{KWPRL0493} in practical realization by using two conditional phase shifts $\theta$ with opposite signs based on weak cross-Kerr nonlinearity because $\theta$ is proportional to $\chi^{(3)}$, with $\chi^{(3)}$ being the third-order nonlinear susceptibility that cannot be changed easily. As illustrated in Ref.~\cite{PPRA0877}: ``{\it Unfortunately, it is generally not possible to change the sign of the conditional phase shift. The nonlinear susceptibility $\chi^{(3)}$ is a material constant that cannot readily be changed. Using different materials with opposite nonlinearities seems highly impractical.}'', proposed by Pieter Kok. Therefore, the required nonlinearity strengths are difficult to achieve in experiment. Another way to produce conditional phase shifts $\theta$ and $-\theta$ is to use giant cross-Kerr nonlinearity. However, it is extremely challenging to achieve giant cross-Kerr nonlinearity under the current experimental conditions.

To enable strong nonlinear interactions between single photons effectively, a lot of schemes, which are based on hybrid (photon-matter) systems that hold great promise for quantum information processing since they allow exploiting different quantum systems at the best of their potentials, have been proposed by using atoms~\cite{BDLCN04428, LHPRL0492, BHTJMDGPRL09102, JMCPRA0979, JJMGJPB1043}, ions~\cite{RDN08453, LCPRA0572}, semiconductor QDs~\cite{CAJWJPRB0878, CWJPRB0878}, and superconducting circuits~\cite{ADALRJSSRN04431} as local matter qubits. Recently,  based on giant optical circular birefringence induced by a single QD spin in a double-sided optical microcavity, Hu {\it et al.}~\cite{CWJJPRB0980} proposed an entanglement beam splitter scheme in which the photon-spin, photon-photon, and spin-spin entanglements can be deterministically generated with high fidelity and high efficiency. Bonato {\it et al.}~\cite{CFSJDMDPRL10104} showed that a single-electron-charged QD in the weak-coupling cavity QED regime exhibited a good interaction between a photon and an electron spin and the hybrid entanglement and CNOT gate between a photon and an electron spin could be efficiently realized via this spin-cavity unit. We also proposed a scheme for optically controlled phase gate and teleportation of a CNOT gate for spin qubits in quantum dot-microcavity coupled system~\cite{HASKPRA1387}. To implement the nonlocal CNOT between two remote electron spins, an entangled Einstein-Podolsky-Rosen (EPR) photon pairs were consumed. In this paper, assisted with a quantum-dot spin in a double-sided optical microcavity, we propose a scalable scheme for constructing a local two-qubit CNOT gate and realizing entanglement swapping for photonic qubits based on spin selective photon reflection from the cavity. The scheme is deterministic and can be
achieved in a nondestructive and heralded way. The feasibility of the scheme is assessed showing that high average fidelities of the CNOT gate and entanglement swapping are achievable in both the weak coupling and the strong coupling regimes. The proposed scheme is simple and feasible as only single-spin rotation and single-spin measurement are required.

The paper is organized as follows. In Sec.~II, we describe the basic building model of QD-microcavity coupled system. In Sec.~III, we show how to construct a deterministic photonic CNOT gate in a nondestructive way. In Sec.~IV, we illustrate the realization of entanglement swapping between photonic qubits with the probability of success $100\%$ in principle. In Sec.~V, we analyze and discuss the experimental challenge for the present scheme. A conclusion is given in Sec.~VI.

\begin{figure}
\includegraphics[width=4.0in]{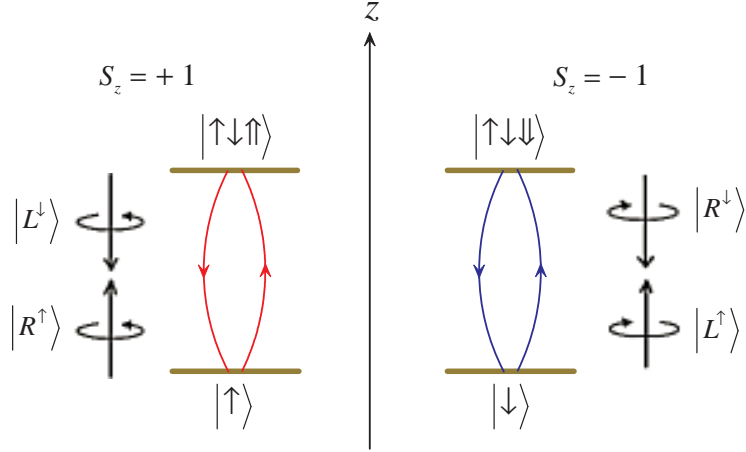}\caption{Relevant energy level and optical selection rules for the optical
transition of $X^-$. Here $|\Uparrow\rangle=|\frac{3}{2},\frac{3}{2}\rangle$ and $|\Downarrow\rangle=|\frac{3}{2},-\frac{3}{2}\rangle$
represent heavy hole states with spin $3/2$ and $-3/2$ components and the superscript arrow is to indicate their propagation direction along the $z$ axis.}
\end{figure}

\section{Cavity-induced photon-spin electron-spin interface }\label{sec1}
Consider a singly charged GaAs/InAs QD, whose relevant energy levels and optical selection rules is shown in Fig.~1, being embedded in a double-sided optical microcavity with both the top and bottom mirrors partially reflective. The optical excitation of the system will produce an exciton ($X^-$) with negative charges and the charged exciton consists of two electrons bound in one hole. According to Pauli's exclusion principle, there are two kinds of optical dipole transitions between the electron and the exciton $X^-$, one involving a $s_z=+1$ photon and the other a $s_z=-1$ photon. If the electron is in the spin-up state $|\uparrow\rangle$, for the photon with $s_z=+1$ ($|R^\uparrow\rangle$ or $|L^\downarrow\rangle$), it feels a coupled (hot) cavity and will be reflected with both the polarization and propagation direction of the photon being flipped. While for the photon with $s_z=-1$ ($|R^\downarrow\rangle$ or $|L^\uparrow\rangle$), it feels an uncoupled (cold) cavity and will be transmitted by the cavity, acquiring a $\pi~{\rm mod}~ 2\pi$ phase shift relative to the reflected photon. In the same way, if the electron is in the spin-down state $|\downarrow\rangle$, the photon with $s_z=+1$ are transmitted and the photon with $s_z=-1$ are reflected by the optical cavity. Therefore, the electron-spin-cavity system behaves like a beam splitter. Based on the transmission and reflection rules of the cavity for an incident circular polarization photon with $s_z=\pm 1$ conditioned
on the QD-spin state, the dynamics of the interaction between photon and electron
in QD-microcavity coupled system is described as below~\cite{CFSJDMDPRL10104}:
\begin{eqnarray}\label{e1}
&&|R^\uparrow, \uparrow\rangle\rightarrow|L^\downarrow, \uparrow\rangle,~~~~~~~~~~~~|L^\uparrow, \uparrow\rangle\rightarrow -|L^\uparrow, \uparrow\rangle,\cr\cr&&|R^\downarrow, \uparrow\rangle\rightarrow -|R^\downarrow, \uparrow\rangle,~~~~~~~~~~|L^\downarrow, \uparrow\rangle\rightarrow |R^\uparrow, \uparrow\rangle,\cr\cr&&|R^\uparrow, \downarrow\rangle\rightarrow -|R^\uparrow, \downarrow\rangle,~~~~~~~~~~|L^\uparrow, \downarrow\rangle\rightarrow |R^\downarrow, \downarrow\rangle,\cr\cr&&|R^\downarrow, \downarrow\rangle\rightarrow |L^\uparrow, \downarrow\rangle,~~~~~~~~~~~~~|L^\downarrow, \downarrow\rangle\rightarrow -|L^\downarrow, \downarrow\rangle,
\end{eqnarray}
where $|L\rangle$ and $|R\rangle$ represent the states of the left- and right-circularly-polarized photons, respectively. The superscript arrow in the photon state indicates the propagation direction along the $z$ axis, and the arrows denote the direction of the electrons.

In a realistic $X^-$-cavity system, the reflection and transmission coefficients of the coupled and the uncoupled cavities are generally different when the side leakage and cavity loss are taken into account. The reflection and transmission coefficients of a double-sided optical microcavity system in the weak excitation approximation are described by~\cite{CWJJPRB0980}
\begin{eqnarray}\label{e2}
r(\omega)&=&\frac{\left[i(\omega_{X^-}-\omega)+\frac{\gamma}{2}\right]\left[i(\omega_c-\omega)+\frac{\kappa_s}{2}\right]+g^2}{\left[i(\omega_{X^-}-\omega)+\frac{\gamma}{2}\right]\left[i(\omega_c-\omega)+\kappa+\frac{\kappa_s}{2}\right]+g^2},\cr\cr
t(\omega)&=&\frac{-\kappa\left[i(\omega_{X^-}-\omega)+\frac{\gamma}{2}\right]}{\left[i(\omega_{X^-}-\omega)+\frac{\gamma}{2}\right]\left[i(\omega_c-\omega)+\kappa+\frac{\kappa_s}{2}\right]+g^2},
\end{eqnarray}
where $g$ is the coupling strength, $\kappa$, $\kappa_s$, and $\gamma$ are the cavity field decay rate, leaky rate, and $X^-$ dipole decay rate, respectively. $\omega$, $\omega_c$, and $\omega_{X^-}$ are the frequencies of the input photon, cavity mode, and the spin-dependent optical transition, respectively. By setting $\omega_c=\omega_{X^-}=\omega$, the reflection and transmission coefficients of the coupled cavity and uncoupled cavity ($g=0$) are given by
\begin{eqnarray}\label{e3}
r(\omega)=\frac{\gamma \kappa_s+4g^2}{\gamma(2\kappa+\kappa_s)+4g^2},~~~~~
t(\omega)=-\frac{2\gamma\kappa}{\gamma(2\kappa+\kappa_s)+4g^2},
\end{eqnarray}
and
\begin{eqnarray}\label{e4}
r_0(\omega)=\frac{\kappa_s}{2\kappa+\kappa_s},~~~~~~~~~~~~~~~~
t_0(\omega)=-\frac{2\kappa}{2\kappa+\kappa_s}.~~~~~~~~~~~~~
\end{eqnarray}
In the following, we investigate how to construct a deterministic CNOT gate and implement entanglement swapping between photonic qubits based on spin selective photon reflection from the optical microcavity discussed above.

\begin{figure}
\includegraphics[width=4.3in]{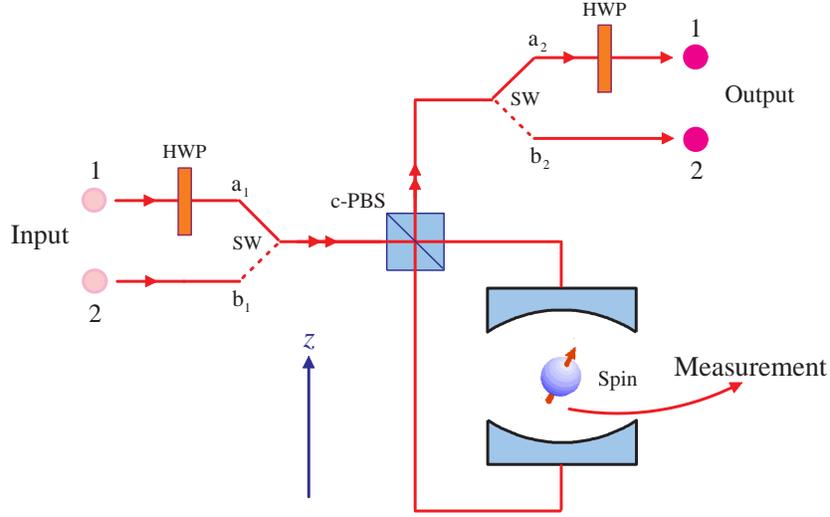}\caption{Schematic of deterministic photonic CNOT gate sing a QD spin in a double-sided optical microcavity. $c$-PBS denotes the polarizing beam splitter in the circular
basis, HWP is half-wave plate, and SW is optical switch.}
\end{figure}

\section{Nondestructive implementation of deterministic CNOT gate for photonic qubits}\label{sec2}
Consider two photons 1 and 2 initially prepared in the polarization states $|\varPsi^{\rm ph}\rangle_1=\alpha|R\rangle_1+\beta|L\rangle_1$ and $|\varPsi^{\rm ph}\rangle_2=\delta|R\rangle_2+\gamma|L\rangle_2$, and an electron spin in the state $|\varPsi^{\rm s}\rangle=(|\uparrow\rangle_s-|\downarrow\rangle_s)/\sqrt{2}$. The schematic of deterministic photonic CNOT gate, with photon 1 being the control qubit and photon 2 being the target qubit, is shown in Fig.~2. The two photons 1 and 2 come in succession to the optical microcavity. The action of the half-wave plate (HWP) is given by the transformation
\begin{eqnarray}\label{e5}
|R\rangle\rightarrow\frac{1}{\sqrt{2}}(|R\rangle+|L\rangle),\cr\cr
|L\rangle\rightarrow\frac{1}{\sqrt{2}}(|R\rangle-|L\rangle).
\end{eqnarray}
The optical switch (SW) controls the photons 1 and 2 in turn making that photon 1 first pass through the optical microcavity and then photon 2. The time interval $\Delta t$ between photons 1 and 2 should be less than the spin coherence time $T^e$. $c$-PBS is polarizing beam splitter in the circular basis, which transmits the input right-circularly-polarized photon $|R\rangle$ and reflects the left-circularly-polarized photon $|L\rangle$. In Fig.~2, before the photon 2 is injected into the spin-cavity system, a Hadamard gate operation, which can be achieved by a $\pi/2$ microwave pulse or optical pulse~\cite{CAJWJPRB0878, DTBYN08456}, is performed on electron spin to accomplish the transformation
\begin{eqnarray}\label{e6}
|\uparrow\rangle_s\rightarrow\frac{1}{\sqrt{2}}(|\uparrow\rangle_s+|\downarrow\rangle_s),\cr\cr
|\downarrow\rangle_s\rightarrow\frac{1}{\sqrt{2}}(|\uparrow\rangle_s-|\downarrow\rangle_s).
\end{eqnarray}
After the photon 2 passes through the optical microcavity, the electron spin is rotated by the Hadamard gate transformation again. Finally, at the output port, the total state of two photons with one spin is transformed into
\begin{eqnarray}\label{e7}
|\varPsi^{\rm ph}\rangle_1\otimes|\varPsi^{\rm ph}\rangle_2\otimes|\varPsi^{\rm s}\rangle\longrightarrow~~~~~~~~~~~~~~~~~~~~~~~~~~~~~~~~~~~~~~~~~~~~~~~~~~~~~~~~~~~~~\cr\cr
(\alpha\delta|R\rangle_1|R\rangle_2+\alpha\gamma|R\rangle_1|L\rangle_2-\beta\delta|L\rangle_1|L\rangle_2-\beta\gamma|L\rangle_1|R\rangle_2)|\uparrow\rangle_s~
\cr\cr+(\alpha\delta|R\rangle_1|R\rangle_2+\alpha\gamma|R\rangle_1|L\rangle_2+\beta\delta|L\rangle_1|L\rangle_2+\beta\gamma|L\rangle_1|R\rangle_2)|\downarrow\rangle_s.
\end{eqnarray}
After the measurement performed on the electron spin, the deterministic CNOT gate between photons 1 and 2, which is unequivocally associated to the measurement results of the electron spin in the $|\uparrow\rangle_s$, $|\downarrow\rangle_s$ basis, is achieved in a nondestructive way (see Table I).

\begin{table}
\caption{The correspondence between the spin measurement results and the state of photons 1 and 2, and the corresponding single-qubit operations on photons 1 and 2 in the case of the deterministic CNOT gate.}\label{0}
\begin{tabular}{ccc}\hline\hline
{~~~~~~~Spin~~~~~~~}& {~~~~~~~~~~~~~~~~~~~~~~~~~~~~Photons 1 and 2~~~~~~~~~~~~~~~~~~~~~~~~~~~~}  &{~~~~~~~Operations~~~~~~~}\\\hline
$|\uparrow\rangle_s$&$\alpha|R\rangle_1(\delta|R\rangle_2+\gamma|L\rangle_2)-\beta|L\rangle_1(\delta|L\rangle_2+\gamma|R\rangle_2)$&
$\sigma_z^1\otimes I^2$\\
$|\downarrow\rangle_s$&$\alpha|R\rangle_1(\delta|R\rangle_2+\gamma|L\rangle_2)+\beta|L\rangle_1(\delta|L\rangle_2+\gamma|R\rangle_2)$&
$I^1\otimes I^2$\\\hline\hline
\end{tabular}
\end{table}

\begin{figure}
\includegraphics[width=5.3in]{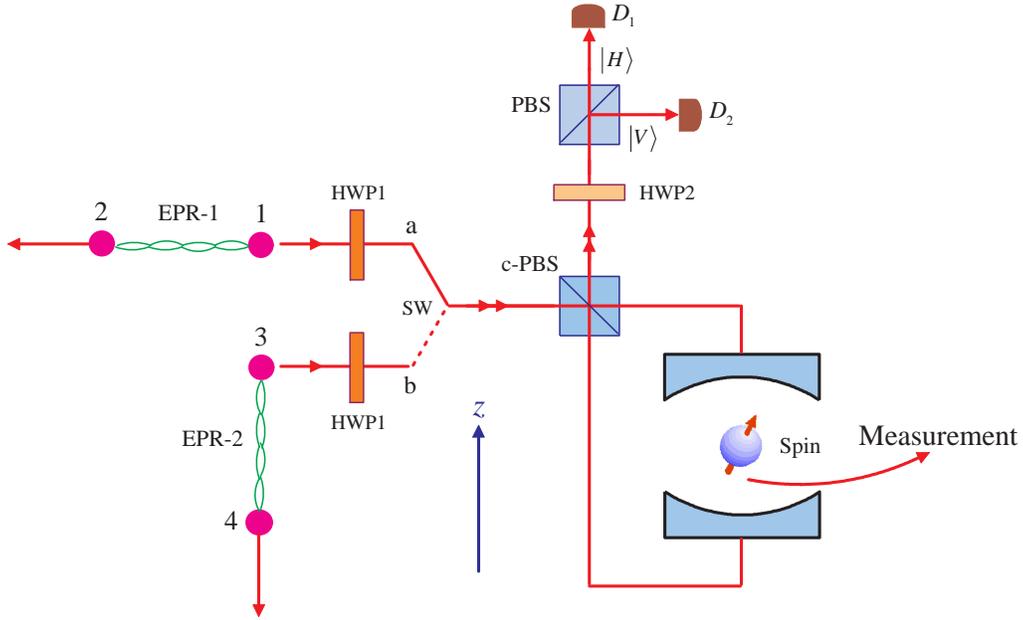}\caption{Schematic of entanglement swapping between photonic qubits.}
\end{figure}

\section{Entanglement swapping between two EPR photon pairs}\label{sec3}
In this section we show how to realize entanglement swapping between two EPR photon pairs. The essential feature of entanglement swapping is that: Given two pairs of entangled systems, $a$, $b$ and $a^\prime$, $b^\prime$, it is possible to generate entanglement between systems $a$, $a^\prime$ ($b$, $b^\prime$)
by a suitable joint measurement on systems $b$, $b^\prime$ ($a$, $a^\prime$).  The state of the systems $a$, $a^\prime$ ($b$, $b^\prime$) is unknown and the state is absolutely random in one of the four EPR pairs with the certain probability before the joint measurement on systems $b$, $b^\prime$ ($a$, $a^\prime$) are made.
The schematic of entanglement swapping between two independent EPR photon pairs is shown in Fig.~3. Photons 1 and 2 are prepared in the state $|\varPsi^{\rm ph}\rangle_{12}^\prime=(|R\rangle_1|R\rangle_2+|L\rangle_1|L\rangle_2)/\sqrt{2}$, photons 3 and 4 in the state
$|\varPsi^{\rm ph}\rangle_{34}^\prime=(|R\rangle_3|R\rangle_4+|L\rangle_3|L\rangle_4)/\sqrt{2}$. The electron spin is initialized to the state $|\varPsi^{\rm s}\rangle^\prime=(|\uparrow\rangle_s+|\downarrow\rangle_s)/\sqrt{2}$. Photons 1 and 3 are then targeted to the spin-cavity system and they pass through the optical microcavity one after another. The action of the half-wave plate, HWP1, is to transform the states as $|R\rangle\rightarrow(|R\rangle+|L\rangle)/\sqrt{2}$ and $|L\rangle\rightarrow(|R\rangle-|L\rangle)/\sqrt{2}$, and HWP2 is used to complete the transformation between the linear polarization and the circular polarization, $|R\rangle\leftrightarrow|H\rangle$ and $|L\rangle\leftrightarrow|V\rangle$. After the reflection and transmission of photons 1 and 3 from the optical microcavity in succession, the total state of one spin with four photons is transformed into
\begin{eqnarray}\label{e8}
|\varPsi^{\rm ph}\rangle_{12}^\prime\otimes|\varPsi^{\rm ph}\rangle_{34}^\prime\otimes|\varPsi^{\rm s}\rangle^\prime\longrightarrow~~~~~~~~~~~~~~~~~~~~~~~~~~~~~~~~~~~~~~~~~~~~~~~~~~~\cr\cr
\frac{1}{4}{\big[}(|H\rangle_1|H\rangle_3-|V\rangle_1|V\rangle_3)|-\rangle_s(|R\rangle_2|L\rangle_4+|L\rangle_2|R\rangle_4)~~~
\cr\cr+(|H\rangle_1|H\rangle_3+|V\rangle_1|V\rangle_3)|+\rangle_s(|R\rangle_2|R\rangle_4+|L\rangle_2|L\rangle_4)~~~
\cr\cr+(|H\rangle_1|V\rangle_3+|V\rangle_1|H\rangle_3)|+\rangle_s(|R\rangle_2|R\rangle_4-|L\rangle_2|L\rangle_4)~~~
\cr\cr-(|H\rangle_1|V\rangle_3-|V\rangle_1|H\rangle_3)|-\rangle_s(|R\rangle_2|L\rangle_4-|L\rangle_2|R\rangle_4){\big]},
\end{eqnarray}
where
\begin{eqnarray}\label{e9}
|+\rangle_s=\frac{1}{\sqrt{2}}(|\uparrow\rangle_s+|\downarrow\rangle_s),\cr\cr
|-\rangle_s=\frac{1}{\sqrt{2}}(|\uparrow\rangle_s-|\downarrow\rangle_s).
\end{eqnarray}
After the spin and photon (1,3) polarization measurements, photons 2 and 4 get entangled in the four EPR pairs, which are unequivocally associated to the measurement results consisting of the spin in the $|+\rangle_s$, $|-\rangle_s$ basis and the polarizations of photons 1 and 3 in the $|H\rangle$, $|V\rangle$ basis, please see Table II in detail.

\begin{table}
\caption{The correspondence between the spin and photons 1, 3 polarization measurement results and photons 2 and 4 in the Bell states in the case of the entanglement swapping.}\label{0}
\begin{tabular}{ccc}\hline\hline
{~~~~~~~~Photons 1 and 3~~~~~~~~}& {~~~~~~~~~~~~~~~~~~~Spin~~~~~~~~~~~~~~~~~~~}  &{~~~~~~~~~~~Photons 2 and 4~~~~~~~~~~~}\\\hline
$|H\rangle_1|H\rangle_3 ~{\rm or}~ |V\rangle_1|V\rangle_3 $&$|+\rangle_s$&
$(|R\rangle_2|R\rangle_4+|L\rangle_2|L\rangle_4)/\sqrt{2}$\\
$|H\rangle_1|V\rangle_3 ~{\rm or}~ |V\rangle_1|H\rangle_3 $&$|+\rangle_s$&
$(|R\rangle_2|R\rangle_4-|L\rangle_2|L\rangle_4)/\sqrt{2}$\\
$|H\rangle_1|H\rangle_3 ~{\rm or}~ |V\rangle_1|V\rangle_3 $&$|-\rangle_s$&
$(|R\rangle_2|L\rangle_4+|L\rangle_2|R\rangle_4)/\sqrt{2}$\\
$|H\rangle_1|V\rangle_3 ~{\rm or}~ |V\rangle_1|H\rangle_3 $&$|-\rangle_s$&
$(|R\rangle_2|L\rangle_4-|L\rangle_2|R\rangle_4)/\sqrt{2}$\\\hline\hline
\end{tabular}
\end{table}

\section{Analysis and discussion}\label{sec4}
The key component in our scheme for the deterministic CNOT gate and entanglement swapping is the
spin-cavity unit. Under the ideal case, the probability of success of our scheme is 100\% in principle. For a realistic spin-cavity unit, however, the side leakage and cavity loss cannot be neglected and the transmission part $t(\omega)$ represented by Eq.~({\ref {e3}}) in the coupled cavity and the reflection part $r_0(\omega)$ represented by Eq.~({\ref {e4}}) in the uncoupled cavity will induce bit-flip errors, which will affect the fidelities of the CNOT gate and entanglement swapping. In this case the rules for optical transitions $X^-$ in a realistic spin-cavity unit is given by~\cite{CWJJPRB0980}
\begin{figure}
\centering
\includegraphics[width=3in]{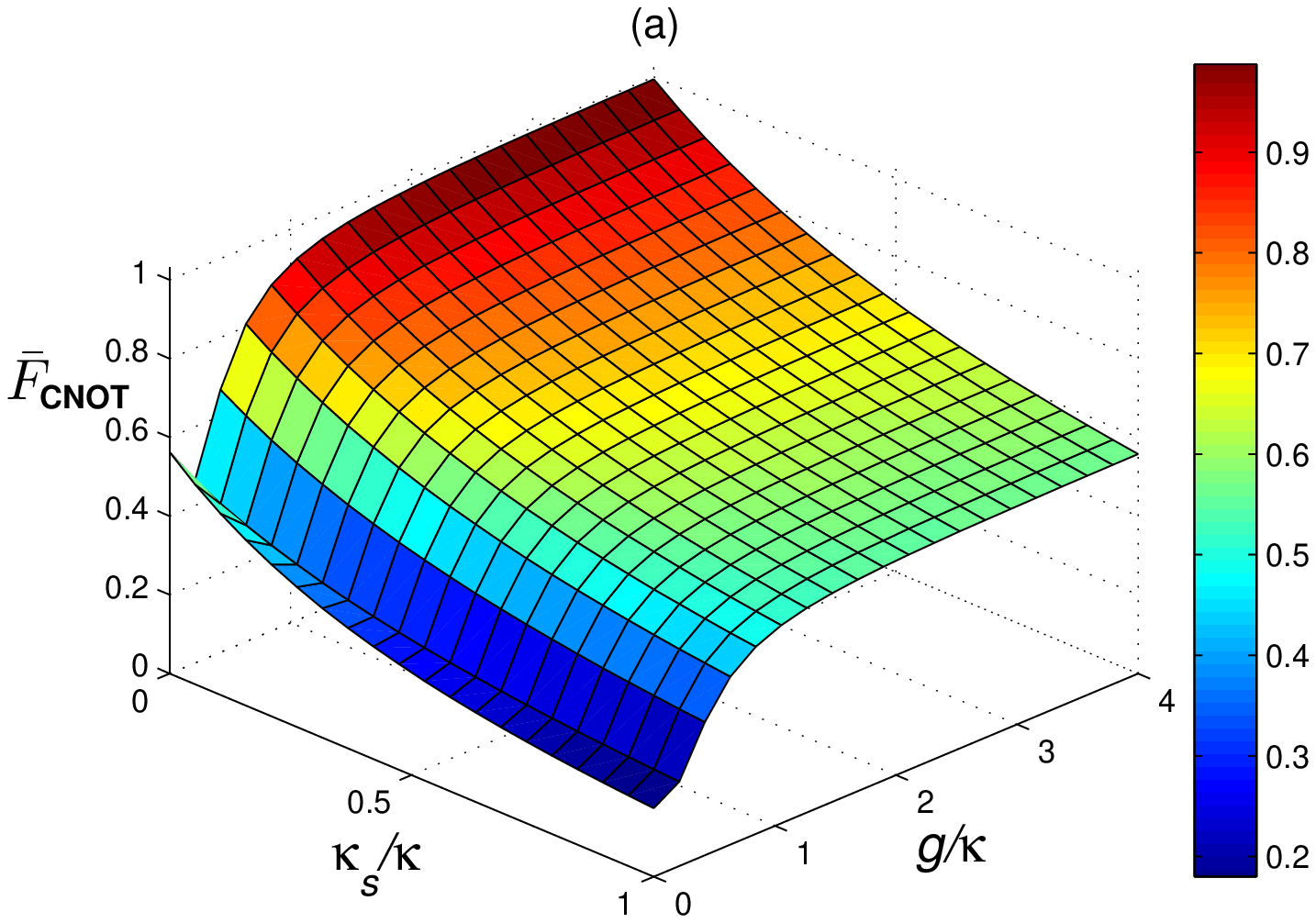}%
\hspace{0.3in}%
\includegraphics[width=3in]{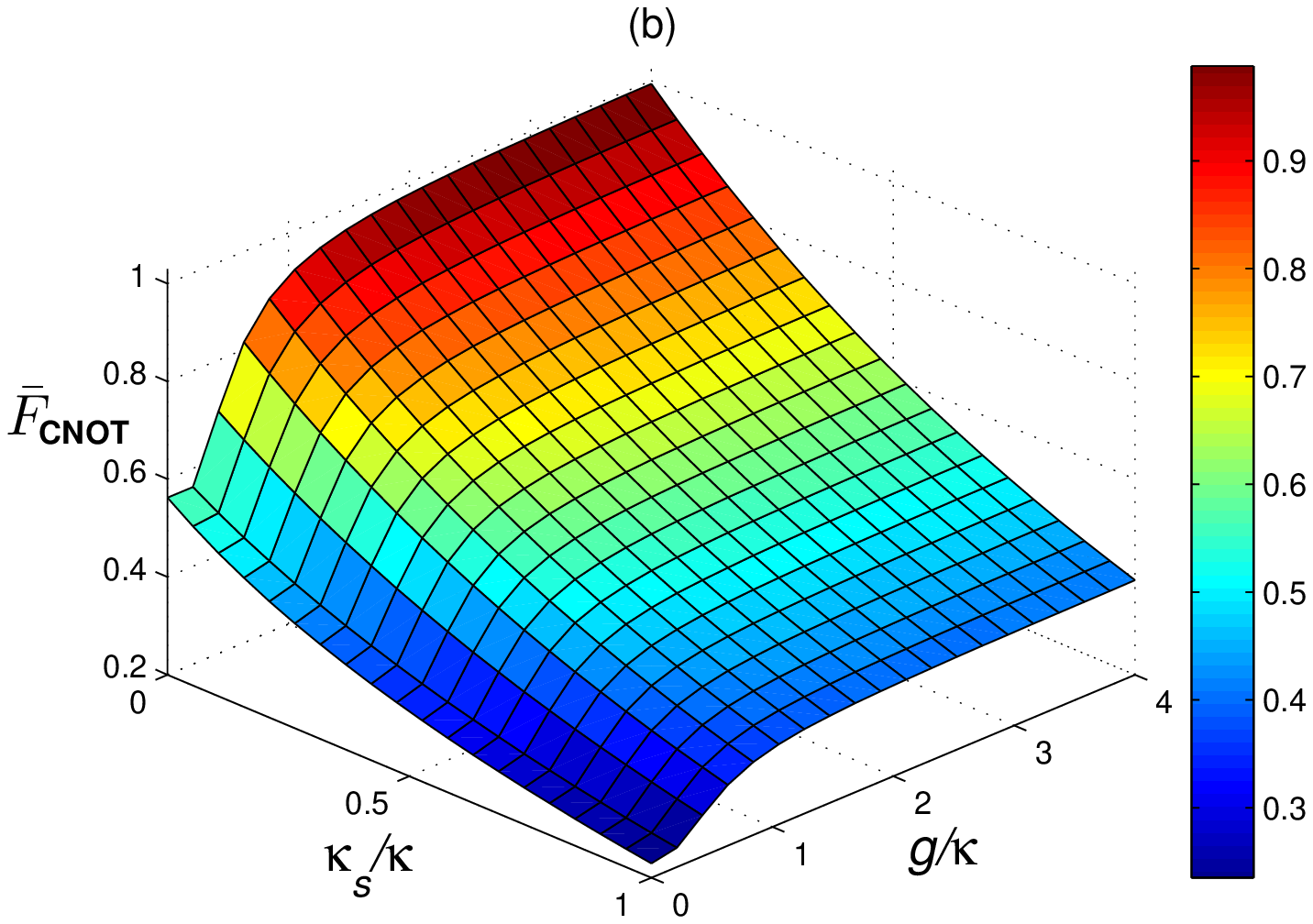} \caption{The average fidelity of the photonic CNOT gate versus the normalized coupling strengths $\kappa_s/\kappa$ and $g/\kappa$. (a) The fidelity corresponds to that the measurement result of the electron spin is $|\uparrow\rangle_s$. (b) The fidelity corresponds to that the measurement result of the electron spin is $|\downarrow\rangle_s$. Here we have set $\gamma=0.1\kappa$.}
\end{figure}
\begin{eqnarray}\label{e10}
|R^\downarrow,\uparrow\rangle&\rightarrow& -|t_0(\omega)||R^\downarrow,\uparrow\rangle-|r_0(\omega)||L^\uparrow,\uparrow\rangle,\cr\cr
|L^\uparrow,\uparrow\rangle&\rightarrow& -|t_0(\omega)||L^\uparrow,\uparrow\rangle-|r_0(\omega)||R^\downarrow,\uparrow\rangle,\cr\cr
|R^\downarrow,\downarrow\rangle&\rightarrow& |r(\omega)||L^\uparrow,\downarrow\rangle+|t(\omega)||R^\downarrow,\downarrow\rangle,\cr\cr
|L^\uparrow,\downarrow\rangle&\rightarrow& |r(\omega)||R^\downarrow,\downarrow\rangle+|t(\omega)||L^\uparrow,\downarrow\rangle.
\end{eqnarray}
To qualify the performance of the CNOT gate, we introduce the
gate fidelity defined as~\cite{JJPPRL9778}
\begin{eqnarray}\label{e11}
\mathcal {F}_{\rm CNOT}=\overline{\langle\varPsi_0|U^\dag \rho_t U|\varPsi_0\rangle},
\end{eqnarray}
where the overline indicates average over all possible input states $|\varPsi_0\rangle$, $U$ is the ideal two-qubit
CNOT gate, and $\rho_t=|\varPsi_t\rangle\langle \varPsi_t|$, with $|\varPsi_t\rangle$ being the final
state after the CNOT gate operation in our scheme. Assume that the two photons are initially in the
general state $|\varPsi_0\rangle=(\cos\theta_1|R\rangle_1+\sin\theta_1|L\rangle_1)\otimes
(\cos\theta_2|R\rangle_2+\sin\theta_2|L\rangle_2)$, the ideal
target state is $|\varPsi_s\rangle=[\cos\theta_1|R\rangle_1\otimes(\cos\theta_2|R\rangle_2+\sin\theta_2|L\rangle_2)
+\sin\theta_1|L\rangle_1\otimes
(\cos\theta_2|L\rangle_2+\sin\theta_2|R\rangle_2)]$.
Then the average fidelity of the CNOT gate can be written as
\begin{eqnarray}\label{e12}
\mathcal {\overline{F}}_{\rm CNOT}=\frac{1}{4\pi^2}\int_0^{2\pi}{\rm d}\theta_1\int_0^{2\pi}{\rm d}\theta_2|\langle \varPsi_s|\varPsi_t\rangle|^2.
\end{eqnarray}
For entanglement swapping, the fidelity is defined as $\mathcal {\overline{F}}_{\rm ES}=|\langle\varPsi_i|\varPsi_r\rangle|^2$,
where $|\varPsi_i\rangle$ represented by Eq.~({\ref {e8}}) and $|\varPsi_r\rangle$ are the final states of the one electron spin and four photons in our scheme for entanglement swapping in the ideal condition and the realistic condition, respectively. We plot the fidelities $\mathcal {\overline{F}}_{\rm CNOT}$ and $\mathcal {\overline{F}}_{\rm ES}$ with the side leakage $\kappa_s/\kappa$ and the coupling strength $g/\kappa$, as shown in Fig.~4 and Fig.~5, which show that the cavity side leakage and cavity field decay have a great impact on the gate fidelity with the increase of leaky rate $\kappa_s$ and decay rate $\kappa$. Here we have assumed that the measurements on electron spin and photon, the single-qubit Hadamard  gate operation on electron spin, and the linear optical elements, such as HWP, $c$-PBS, PBS, and optical switches, are all perfect. Recently, much effort has been made on optical spin
cooling~\cite{MJAAKAS06312} and optical spin manipulating~\cite{DTBYN08456, JMNLDS08320} in QDs, which provide a effective method for the spin state measurement. As reported in Refs.~\cite{DTBYN08456, JMNLDS08320}, the superposition state of spin can be realized from the eigenstates by performing single spin rotations with nanosecond ESR pulses or picosecond optical pulses and the quantum Zeno effect could be used to maintain the prepared states. Moreover, linear optical method is a promising approach for constructing quantum networks. Tunable linear optical elements can be made using polarizers and polarizing beam splitters. The computational power of passive and active linear optical elements have been investigated showing that linear optical elements are enough to implement reliable quantum computation and provide important network primitives such as multiplexing and routing~\cite{ERGQUAN0006088, JPPRA1286}. These imply the feasibility of our scheme using QDs in the microcavity system.

\begin{figure}
\includegraphics[width=4.0in]{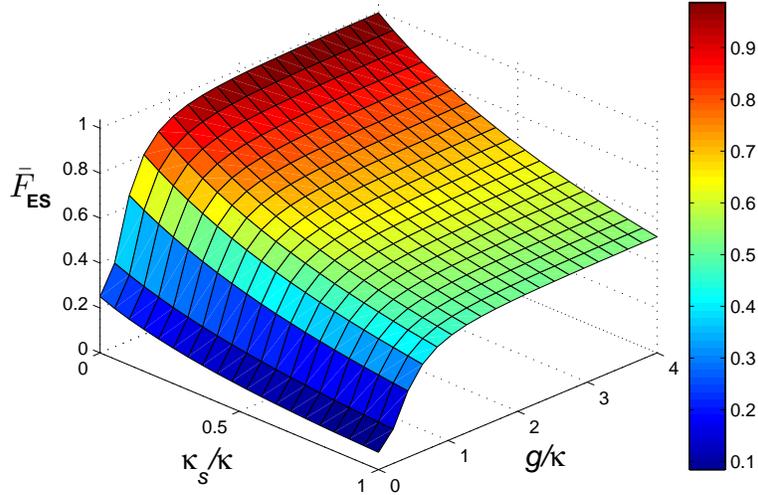}\caption{The average fidelity of the entanglement swapping versus the normalized coupling strengths $\kappa_s/\kappa$ and $g/\kappa$, where we have set $\gamma=0.1\kappa$.}
\end{figure}

Experimentally, the weak coupling with $g<(\kappa+\kappa_s)/4$ can be easily achieved, while for strong coupling with $g>(\kappa+\kappa_s)/4$, which is still a big challenge, has also been observed in QD spin-cavity systems~\cite{JGACSSLVTAN04432, TAJGHGCODN04432, EPDAJJJPRL0595, SCAMSCASMAAPL0790}, and $g/(\kappa+\kappa_s)\simeq 0.5$ and $g/(\kappa+\kappa_s)\simeq 2.4$ were reported~\cite{JGACSSLVTAN04432, SCAMSCASMAAPL0790}. In our scheme, if setting $\kappa_s=0.5\kappa$, $g=2.5\kappa$, we can obtain $\mathcal {\overline{F}}_{\rm CNOT}=93.74\%$ and $\mathcal {\overline{F}}_{\rm ES}=93.71\%$; even when setting $\kappa_s=1.0\kappa$, $g=0.45\kappa$, we also can obtain $\mathcal {\overline{F}}_{\rm CNOT}=71.09\%$ and $\mathcal {\overline{F}}_{\rm ES}=70.34\%$. Therefore, our scheme can work in both the weak coupling and the strong coupling regimes. Furthermore, the fidelities may be reduced by the following two factors~\cite{CJPRB1183}
\begin{eqnarray}\label{e13}
F_1^\prime&=&\frac{1}{2}\left[1+e^{-\frac{\Delta t}{T_e}}\right],\cr\cr
F_2^\prime&=&1-e^{-\frac{\tau}{T_c}},
\end{eqnarray}
due to the spin decoherence and the trion dephasing caused by the exciton decoherence, respectively.
Here $T_e$ and $\Delta t$ are the electron spin coherence time and the time interval between two input photons for the CNOT gate and entanglement swapping, respectively; $\tau$ and $T_c$ are the cavity photon lifetime and the exciton coherence time, respectively. To get high fidelities, the time interval $\Delta t$, which limited by the critical photon and the cavity photon lifetime, should be shorter than the spin coherence time $T_e$, i.e., $\Delta t<T_e$. The time interval between two input photons is defined as $\Delta t=\tau/n_0$, with $n_0=\gamma^2/2g^2$ being the number of the critical photons. For a micropillar microcavity  with $d=1.5~\mu{\rm m}$ and $Q=1.7\times 10^4$, the critical photon number $n_0=2\times10^{-3}$ and the cavity photon lifetime $\tau=9~{\rm ps}$ can be achieved by taking $g/(\kappa+\kappa_s)=1.0$, $\kappa_s/\kappa=0.7$, and $\gamma/\kappa=0.1$~\cite{CJPRB1183}. Therefore, the time interval between two input photons could be longer than $\tau/n_0=4.5~{\rm ns}$, which is several orders shorter than the single electron charged QD spin choherence time $T_e\simeq 3~\mu{\rm s}$~\cite{JAJEAMCMAS05309, ADAAIRVDAMS06313}.
On the other hand, the trion dephasing, including the optical dephasing and the spin dephasing, should be considered here. In self-assembled In(Ga)As QDs, the optical coherence time of excitons, which can approach several hundred picoseconds~\cite{PWSURDDPRL0187, DKJPRL0187, WPUVDAPRB0470}, is ten times longer than the cavity photon lifetime (about tens of picoseconds). Therefore, the optical dephasing can only affect the fidelities slightly. Recently, it has been reported that the exciton coherence time $T_c>100~{\rm ns}$ can be achieved~\cite{DBPGKNPRS09325}, which is at least three orders of magnitude longer than the cavity photon lifetime, showing that the spin dephasing can be safely neglected in our considerations. Finally, significant progress has recently been made in the manipulation of single electron spins in QDs~\cite{JPNDS01292, ASSDDATMNP095}. Schemes for fast initialization of the spin state of an electron and optically controlled single-qubit rotations for the spin of an electron in QDs have been demonstrated detailedly in Refs.~\cite{CXDSLPRL0798, XYBQJDADCLPRL0799, DSSMAMTDPRL08101, EKXBDADLPRL10104, CLJPCM0719}. Therefore, the required preparation of electron spin superpositions and single-qubit gate rotation operations on the electron spin in our scheme could be effectively implemented.

\section{Conclusions}\label{sec5}
In conclusion, we have proposed an effective scheme for deterministic construction of a two-qubit CNOT gate and realization of entanglement swapping between photonic qubits using a quantum-dot spin in a double-sided optical microcavity. The CNOT gate and entanglement swapping are nondestructive and heralded by the sequential measurement of electron spin and the single-qubit operations on the photons. The calculated results show that the cavity leakage ($\kappa_s$) and the cavity loss ($\kappa$) greatly affect the fidelity of the scheme. However, high average fidelities of the CNOT gate and entanglement swapping are still achievable in both the weak coupling and the strong coupling regimes. When $\kappa_s\ll \kappa$, the average fidelities of of the CNOT gate and entanglement swapping can approach near unity in the strong coupling regime. The scheme might be experimentally feasible with current technology and would open promising perspectives for long-distance quantum communication and distributed quantum information processing with photons.

\begin{center}
{\small {\bf ACKNOWLEDGMENTS}}
\end{center}

This work is supported by the National Natural Science Foundation
of China under Grant Nos. 11264042, 61068001, and 11165015; the China Postdoctoral
Science Foundation under Grant No. 2012M520612; the Program for Chun Miao Excellent Talents of Jilin Provincial Department of Education under Grant No. 201316; and the Talent Program of Yanbian University of China under Grant No. 950010001.

\end{document}